\documentclass[runningheads]{llncs}

\usepackage[utf8]{inputenc}
\usepackage[T1]{fontenc}
\usepackage{subcaption}
\usepackage{lipsum}
\usepackage{graphicx}
\usepackage{subcaption}
\usepackage[export]{adjustbox}
\usepackage{textcomp}
\usepackage[hang,flushmargin]{footmisc}
\usepackage{url}
\usepackage{import}
\usepackage{color}

\usepackage{soul}
\usepackage{pifont}
\usepackage{enumitem}
\usepackage{multirow}
\usepackage{blindtext}
\usepackage{dirtytalk} 
\usepackage{tabularx} 
\usepackage{multirow}
\usepackage{listings} 
\usepackage{todonotes}
\usepackage{mathtools}
\usepackage{cleveref}
\usepackage{rotating}
\usepackage[hang,flushmargin]{footmisc} 
\usepackage{arydshln}
\usepackage{flushend}

\graphicspath{ {./Figures/} }

\lstdefinestyle{python}{ 
    xleftmargin=6.0ex,
    xrightmargin=2.0ex,
    numbers=left,
    frame=single
}
\definecolor{gray10}{gray}{.9}
\definecolor{arsenic}{rgb}{0.23, 0.27, 0.29}
\usepackage{color}

\RequirePackage{fontawesome}

\newcommand{\keyfindings}[1]{ %
	\vspace{5pt} %
	\noindent\fcolorbox{arsenic}{gray10}{%
		\parbox{0.97\linewidth}{%
			\textbf{\faKey\ } #1 %
		}%
	}%
	\vspace{5pt} %
}%

\definecolor{gray50}{gray}{.5}
\definecolor{gray40}{gray}{.6}
\definecolor{gray30}{gray}{.7}
\definecolor{gray20}{gray}{.8}
\definecolor{gray10}{gray}{.9}
\definecolor{gray05}{gray}{.95}

\newcolumntype{L}[1]{>{\raggedright\let\newline\\\arraybackslash\hspace{0pt}}p{#1}}
\newcolumntype{C}[1]{>{\centering\let\newline\\\arraybackslash\hspace{0pt}}p{#1}}
\newcolumntype{R}[1]{>{\raggedleft\let\newline\\\arraybackslash\hspace{0pt}}p{#1}}

\newlength\Linewidth
\def\findlength{\setlength\Linewidth\linewidth
    \addtolength\Linewidth{-4\fboxrule}
    \addtolength\Linewidth{-3\fboxsep}
}

\begin{document}


\title{One Microservice per Developer: Is This the Trend in OSS? 
}

\author{Dario Amoroso d'Aragona*\inst{1}, Xiaozhou Li*\inst{2}, Tomas Cerny\inst{3}, Andrea Janes\inst{4}, Valentina Lenarduzzi\inst{2}, Davide Taibi\inst{1}\inst{2} }
\authorrunning{Amoroso d'Aragona et al.}

\institute{Tampere University, Finland \email{dario.amorosodaragona@tuni.fi},\\
\and
University of Oulu,Finland \email{xiaozhou.li;valentina.lenarduzzi;davide.taibi@oulu.fi},\\
\and
University of Arizona, USA \email{tcerny@arizona.edu},\\
\and 
Vorarlberg University of Applied Sciences, Austria \email{andrea.janes@fhv.at}\\
}

\maketitle

\begin{abstract}
When developing and managing microservice systems, practitioners suggest that each microservice should be owned by a particular team. In effect, there is only one team with the responsibility to manage a given service. Consequently, one developer should belong to only one team. This practice of "one-microservice-per-developer" is especially prevalent in large projects with an extensive development team.

Based on the bazaar-style software development model of Open Source Projects, in which different programmers, like vendors at a bazaar, offer to help out developing different parts of the system, this article investigates whether we can observe the "one-microservice-per-developer" behavior, a strategy we assume anticipated within microservice based Open Source Projects.

We conducted an empirical study among 38 microservice-based OS projects. Our findings indicate that the strategy is rarely respected by open-source developers except for projects that have dedicated DevOps teams.

\end{abstract}


\section{Introduction}
Microservices are increasing their diffusion both in industry and in Open Source Software (OSS) projects ~\cite{DamianDataSet}. 

Microservices are small and autonomous services deployed independently, with a single and clearly defined purpose~\cite{Fowler2014,NewmanBook2015}. Because of their independent deployment, each microservice can scale independently from others. 
Some authors see microservices not primarily as a technological benefit but also as a way to scale up the number of development teams: "\textit{microservices are not necessarily required to manage huge software, but rather to manage a huge number of people working on them }\cite{Reinfurt2021}." The rationale is that since microservices decouple software components, less communication is necessary to develop them, and larger teams become possible.

Practitioners suggest that a microservice should be \textit{owned} and managed by a single team \cite{DamianDataSet,7436659,Amazon,microservice.io,qcon@2022.darkenergy,Reinfurt2021,Wolff2021}.
The supportive argument sources from ``\texttt{Conway's law}'' \cite{Conway1968} states that ``\textit{organizations which design systems (in the broad sense) are constrained to produce designs which are copies of the communication structures of these organizations}.'' Following this law, it would be ineffective or detrimental to have two separate teams working on one microservice. Working on one microservice requires communication within the team, and if these communication structures are not present, the work on a joint microservice becomes hard. Therefore, it is suggested that each team is responsible for one or more business functions \cite{Amazon,microservice.io}. 
While some authors (and Conway himself \cite{Conway1968}) clearly foresee that each team can own more than one microservice/subsystem \cite{carneiro2016microservices,microservice.io,Wolff2021}, others suggest that "a team should have exactly one service unless there is a proven need to have multiple services", to not exceed the cognitive capacity of a team \cite{microservice.io}. Particularly if business functions are large (e.g., "customer management" or "order management"), practitioners suggest that one team is fully dedicated to~one microservice~\cite{qcon@2022.darkenergy}. 

Therefore, following the practitioners' recommendations~\cite{Conway1968,microservice.io,qcon@2022.darkenergy,Wolff2021,carneiro2016microservices}, a developer must belong only to one team, and each team must contribute only to one microservice. Consequently, we can deduct that each team member, and therefore each developer, must contribute to only one microservice. 
Based on these assumptions, it would be interesting to investigate the  ''one-microservice-per-developer'' strategy to verify to which extent it is  considered~in~practice.



The goal of this paper is to investigate to which degree it is correct to assume that in microservice OSS projects, the one-microservice-per-developer strategy is respected. 

Particularly, teams developing OSS projects using a bazaar-style software development model (as described in Eric Raymond's seminal essay "The Cathedral and the Bazaar," in which different programmers, like vendors at a bazaar, offer to help out developing different parts of the system \cite{Raymond1999}), require a decoupled software architecture, which---what we assume---would manifest in a decoupled collaboration structure. Within OSS projects adopting a microservice architecture, we hypothesize, we observe that developers, during a given time window, commit only to one microservice at a time.

For these aims, we designed and conducted an empirical study among 38 microservice-based OSS projects selected from the dataset created by Baresi et al.~\cite{DamianDataSet}. Using code repositories of these projects and analyzing the history of commits, we calculated the average number of microservices developed by each developer determining how well is the one-microservice-per-developer strategy employed. In addition, we further investigated the potential developer profiles using Exploratory Factor Analysis (EFA) to detect the patterns of developer behaviors in the core contributor groups~of~these~projects.

\textbf{Paper structure:} Section \ref{sec:EmpiricalStudy} describes the empirical study design, while Section \ref{sec:Results} reports the obtained results. Section \ref{sec:Discussion} discusses the results, and Section \ref{sec:ThreatsValidity} highlights the limitation of this work. Section \ref{sec:RelatedWork} presents the related work, and Section \ref{sec:Conclusion} concludes.

\section{Related Work}
\label{sec:RelatedWork}

Developer interaction analysis has been approached from different perspectives in OSS communities. Given the large quantities of produced communication artifacts throughout the developer interaction in the development process, various automated approaches have been proposed. Common sources of input for such analysis include version control systems (performing mining source code repositories) \cite{10.5555/2818754.2818824,10.1145/1984642.1984647}, mailing lists, and issue trackers \cite{6976091,10.1145/1453101.1453107}, or developer online surveys \cite{10.1145/1985793.1985832}.
Prior to the era of microservices, Bird et al. \cite{10.1145/1453101.1453107} considered social network communities and system modularity. They researched code artifact changes across modules and analyzed email archives to derive social networks and assess community alignment with modularity. 
The conclusions and research questions of Bird et al. \cite{10.1145/1453101.1453107} in the scope of microservices drive new perspectives. Microservices are self-contained, and with regard to Conway's law, we can consider well-defined teams assigned to particular microservice development. In addition, the remaining challenge related to crosscutting concerns cannot be simply negated in microservices. 

With regards to microservices and well-defined separation boundaries by code repositories (or at least repository modules). It can thus be assumed that code artifacts modified by developers within the same community are placed in a related repository location.

Throughout OSS software development, it can be expected that developer assignments to subsystems remain stable (i.e., given expertise alignments, subsystem assignment, etc.). Ashraf and Panichella \cite{9461033} analyzed a set of OSS projects to examine developer communities from the perspective of their subsystem assignment and interaction 
highlighting that emerging communities change considerably across a project's lifetime and do align with each other.

The microservices perspective, as suggested by Lenarduzzi et al. \cite{10.1145/3234152.3234191}, enables teams to work independently, reducing cross-team communication. At the same time, upon microservice integration, issues are reported across teams, as suggested by Bogner et al. \cite{10.1007/s10664-021-09999-9} who report on ripple effects. There are other underlying issues behind this relevant to system evolution, such as missing system-centered perspective and lack of tools to analyze coherence across microservices, perform modification trade-off analysis, or evaluate the conformance of the as-built and as-documented architectures.

Besides interaction analysis to understand communities, other interesting research directions took place. For instance, Marco et al. \cite{10.1145/3273934.3273943} analyzed GitHub commit comments regarding emotions and feelings expression showing that ``one-commit'' developers are more active and polite when posting comments as opposed to ``multi-commit'' developers, that are less active in posting comments, and when commenting, they are less polite.

In a timely thesis, Shi \cite{shi2021establishing} looked into establishing contributor roles within software repositories by mining architectural information. 
In a case study on Apache Tomcat, they used the metric to deduce these roles and validate them with particular roles listed on the project website. Such a research direction aligns with the perspective of microservices with established separation of duty. Furthermore, the classification of experts responsible for re-engineering or management can lead to better insights into the applicability of Conway's law across microservice developers.

It is also important to take into account that enterprise companies like Red Hat manage OSS projects \cite{10.1145/3379597.3387495} rather than projects based on volunteer contribution. This can influence role identification, contributor duty spread across modules, and also the community network. Spinellis et al. \cite{10.1145/3379597.3387495} considered the detection of OSS projects that are supported by enterprises. Such projects can serve as better benchmarks for practical case studies.

With respect to inter-project dependency identification, Blincoe et al. \cite{BLINCOE2019174} considered reference coupling. The reference coupling method often identifies technical dependencies between projects that are untracked by developers. Understanding inter-project dependency is important for change impact analysis and coordination. In their study, they manually analyzed identified dependencies and categorized and compared them to dependencies specified by the development team. They also assessed how the ecosystem structure compares with the social behavior of project contributors and owners. As a result, of socio-technical alignment analysis within the GitHub ecosystems, they found that the project owners' social behavior aligns well with the technical dependencies within the ecosystem. Still, the project contributors' social behavior does not align with these dependencies. In microservices, this could possibly translate into system architects aware of consequences and microservice developers who operate in isolation as suggested by Lenarduzzi et al. \cite{10.1145/3234152.3234191} and unaware of such as inter-project dependency.

In a similar perspective, Scaliante Wiese et al. \cite{WIESE2017220} researched co-change prediction. They use issues, developers' communication, and commit metadata to analyze change patterns for prediction models. They demonstrate that such models based on contextual information from software changes are accurate and can support software maintenance and evolution, warning developers when they miss relevant artifacts while performing a software change. 


\section{The Empirical Study}
\label{sec:EmpiricalStudy}

In this section, we describe our empirical study reporting the goal and research questions, context, data collection, and data analysis 
following the guideline defined by Wohlin et al. \cite{wohlin2012experimentation}. 

Our goal is to evaluate to what extent the one-microservice-per-developer strategy, recommended by practitioners~\cite{DamianDataSet,7436659,Amazon,microservice.io,qcon@2022.darkenergy} is respected in OSS projects. To allow verifiability and replicability, we published the raw data in the replication package\footnote{ \url{https://figshare.com/s/6ba4e0063ab04d03d6d6}\label{Package}}.


Then, we formulated two Research Questions (RQs). 


\smallskip
\textbf{RQ$_1$.} How well is the one-microservice-per-developer strategy respected in OSS projects following a microservice architecture?

\smallskip
\textbf{RQ$_2$.} Which developer roles better respect the one-microservice-per-developer strategy?

\smallskip
With \textbf{RQ$_1$}, we investigated if developers are actually responsible, and therefore committing, only to a single microservice. 
In \textbf{RQ$_2$}, we aimed to understand if specific roles are respecting the aforementioned strategy differently. We expect that some roles (e.g. DevOps) can be involved in multiple microservices, while other roles (e.g. coders) are involved only in a single microservice.

\subsection{The Selected Projects}
We considered the manually validated dataset including 145 microservice-based projects, proposed by Baresi et al.~\cite{DamianDataSet}. The authors developed, validated, and released a tool to recognize the architecture (e.g., the microservices, the external services, and the databases used) in a given microservices-based project. In addition, the authors provided a list of 145 projects that have been manually validated as non-toy projects regularly using microservices, for which they also reported the list of built-in microservices in the form of relative paths and some other set-up information not used in our case. In particular, for our analysis, we leveraged the list of projects and the related list of microservices identified by a list of sub-project folders. The dataset consists of projects whose source code is accessible on GitHub\footnote{\url{https://github.com}} and is complemented with further data, including the microservice list.


In order to select a set of relevant projects for our study, we defined the following inclusion criteria: 
\begin{itemize}
    \item Project with at least 2 microservices. With this threshold, we aim to exclude non-microservices projects.
    \item Projects with at least 2 microservices committed in the last 12 months. To analyze projects that are still maintained.
\end{itemize}


By requiring a minimum number of microservices and activities in the last non-representative outliers for our study can be excluded. 
It is important to note that we did not exclude projects based on their  programming languages.

As a result, we included 38 microservice-based projects with a total of 379 microservices (10 microservices per project on average).

\subsection{Data Collection}

To collect statistics about the development process, we  browsed every project commit.
We gathered the timestamp, the author's name, and the precise change locations for each commit. With the latter, a modification is connected to a microservice. We specifically created a heuristic that matches if the path of the modified file is contained in the project's list of microservices. If so, we updated the list of microservice changes in the aforementioned author's commit.

\subsection{Data Analysis}
To answer \textbf{RQ$_1$}, the goal is to investigate the microservice coverage by the developers on average by examining their commits on the microservices. 
Here, we considered only the commits involving source code files and excluded all the commits regarding documentation and setup files. We analyzed the distribution of commits over developers and microservices to understand 1) how many microservices have developers in common, and 2) how many developers work on more than one single microservice.




However, the threat, in this case, is the situation where
a developer finishes work on a microservice and gets started
to work on another microservice, or for some reason, he/she is just moved
to another team of developers. From our point of view, this situation
does not lead to a real violation of the \textit{one-microservice-per-developer} strategy. For this reason, we have defined a metric for counting
how many times a developer recommits to a microservice
after starting work on another microservice; in other words, if a developer
$D_1$ commits to microservice $m_1$, switches the team, and starts
committing to a $m_2$ microservice, then the result of our metric will be $0$ because the developer never goes back to the previous microservice; otherwise, if after a while the developer commits back to $m_1$, our metric results will give $1$, because the developer goes back to the previous microservice ($m_1$).


To answer \textbf{RQ$_2$}, we need to understand how to identify the role of each contributor in OSS microservice projects. Different from industrial projects, within OSS projects on GitHub, contributors, are neither assigned roles by "project managers" or "product owners" nor obliged to focus on the tasks assigned to them in the corresponding areas. Therefore, we shall only be able to understand the roles based on the domains each contributor has been contributing to.

To identify the roles of the project contributors, we adopted the approach proposed by Montandon et al.~\cite{montandon2021mining} combined with the  Exploratory Factor Analysis (EFA). 

Montandon et al.~\cite{montandon2021mining} proposed a machine-learning-based approach  based on the  extensions of the committed files. They used more than 100k developers' data from GitHub together with Stack Overflow data and studied five critical roles: \textit{Backend}, \textit{Frontend}, \textit{Data scientist}, \textit{DevOps}, and \textit{Mobile}. Herein, we initially adopt the same settings. 

The  Exploratory Factor Analysis (EFA) \cite{hair2006multivariate} 
aims to discover not only the number of factors but also what measurable variables together influence which individual factors \cite{decoster1998overview}. With EFA, we can reduce the complexity of the data, and also are able to explain the observations with a smaller set of latent factors. Importantly, by doing so we can also discover the relations among the variables.

Herein, we follow these steps to conduct EFA on our commit dataset and determine the profile of each developer: 

\begin{enumerate}
    \item \textbf{Preprocessing.} Firstly, we group the obtained developer behavior data. 
    \item \textbf{Data Verification.} 
    Secondly, we verify its sampling adequacy and statistical significance. For example, we can use Bartlett's Test of Sphericity \cite{snedecor1989statistical} and Kaiser-Meyer-Olkin (KMO) Test \cite{kaiser1974index} for such a purpose.
    \item \textbf{Determining Factor Number.} 
    Thirdly, we find the number of factors using parallel analysis (PA) \cite{horn1965rationale}. Herein, we employ the Monte Carlo simulation technique to simulate random samples consisting of uncorrelated variables. Then, We extract the eigenvalues of the correlation matrix of the simulated data and compare the extracted eigenvalues that are ordered by magnitude to the average simulated eigenvalues. Significant factors are the ones with ob.served eigenvalues higher than the corresponding simulated eigenvalues
    
    \item \textbf{Factor Extraction and Interpretation.} With the number of factors determined, we conduct the EFA on the dataset. To simplify the interpretation of the factor analysis result, we employ the \textit{varimax} rotation technique \cite{kaiser1958varimax} to maximize the variance of each factor loading. 
    \item \textbf{Determining Individual Developer Role Allocation.} To apply the factor-variable relation to individual contributors, we shall calculate the similarity between the developers' contributions 
    in terms of the languages and each detected factor. By comparing the contributor's similarity to each role factor, we shall understand more intuitively which role(s) he/she leans to. Such results can be visualized in a radar chart. 
    
\end{enumerate}

For this study, as a result of the EFA, we shall have a set of factors, each of which is closely related to a set of latent variables, i.e., programming languages. To be noted, due to the fact that the original data are collected from projects of different programming languages, it is likely that contributors working on different languages lean toward similar roles. For example, contributors working on CSS and VUE can both be \textit{Frontend} contributors. Therefore, we shall observe the loadings of the EFA and manually merge factors related to only closely-connected languages into the unified roles.

Particularly, the contributor's similarity to each role-factor can be calculated using 
the Kumar-Hassebrook (KH) similarity, which incorporates also the inner product of the assigned values of the variables \cite{kumar1990performance}. Moreover, using the KH similarity, we can evaluate each contributor's effort level in each pre-detected role-factor, respectively.

\section{Results}

\label{sec:Results}
In this Section, we report the obtained results to answer our Research Questions (RQs).
\subsection*{$RQ_1$ How well is the one-microservice-per-developer strategy respected in OSS projects following a microservice architecture?}

\begin{figure*}[h]
\centering
\begin{minipage}{.5\textwidth}
  \centering
  \includegraphics[width=\linewidth]{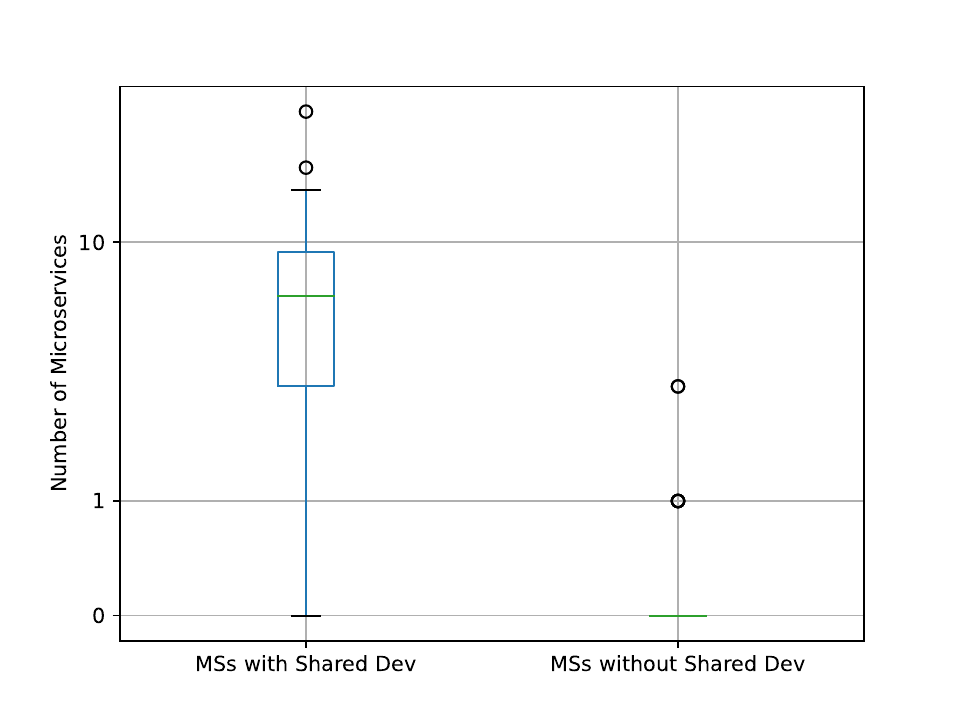}
  \caption{\# microservices with shared and not shared developers (RQ$_1$)} 
  \label{fig:n_ms_shared_not_shared_dev_symlog_scale}
\end{minipage}%
\begin{minipage}{.5\textwidth}
  \centering
  \includegraphics[width=\linewidth]{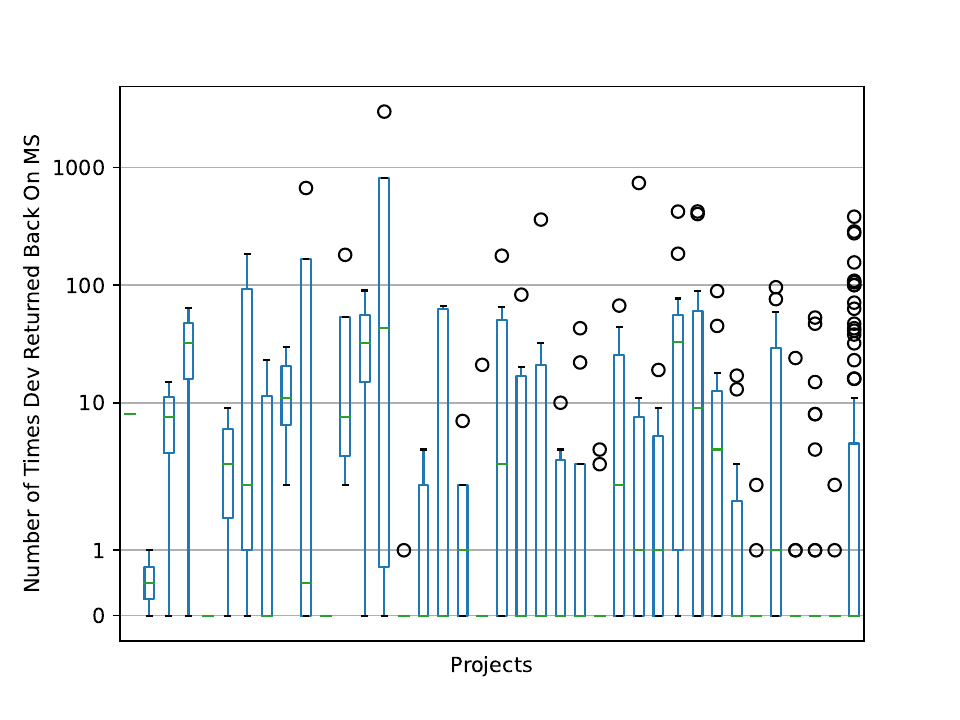}
  \caption{Frequency that developers have committed back (RQ$_1$)} 
  \label{fig:frequency_commit_back_rq2}
\end{minipage}
\label{fig:papers}
\end{figure*}

To answer $RQ_1$, we investigated the single developer, assuming that a single developer does not belong to more than one team at the same time.
\Cref{fig:n_ms_shared_not_shared_dev_symlog_scale} compares the number of microservices with shared developers (\textit{MSs with Shared Dev}) with the number of microservices where all developers committed only to the same microservice  (\textit{MSs without Shared Dev}).

Unexpectedly, only 2 projects always respected the one-microservice-per-developer strategy, while the remaining projects shared among services.

Since the vast majority of the 
projects (\Cref{fig:sortbydv}) a developer works on more than one microservice, we continue our analysis to understand if developers are simply switching teams, or are working on more microservices at the same time.   

\Cref{fig:frequency_commit_back_rq2} shows the result of the number of times developers commit back on a microservice after moving to another one among the projects.  In two projects out of 38, developers never commit back to the previous microservice.
In most of the projects (\textbf{$53$\%}), the median is $0$, in \textbf{$34$\%} of the projects the median is between $1$ and $10$, and finally in the \textbf{$13$\%} of the projects is more than $10$.  However, analyzing the figure, we can see that the boxplots are very stretched, thus in the same project there are some developers that do not return back after changing microservice (or never change microservice) and some developers that instead, commit to the previous microservice.


\keyfindings{
As a result, we conclude that in an OSS context, the \textit{one-microservice-per-developer} strategy is not respected as in most cases, developers work on more than one microservice in parallel.}

\subsection*{$RQ_2$ Which developer roles better respect the one-microservice-per-developer strategy?}

\begin{figure*}[h]
\centering
\begin{minipage}{.5\textwidth}
  \centering
  \includegraphics[width=0.8\linewidth]{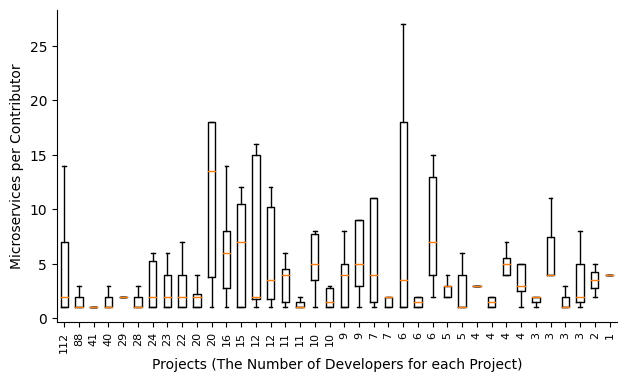}
  \caption{MS Per Developer sorted by \#Developer (RQ$_1$)}
    \label{fig:sortbydv}
\end{minipage}%
\begin{minipage}{.5\textwidth}
  \centering
  \includegraphics[width=0.8\linewidth]{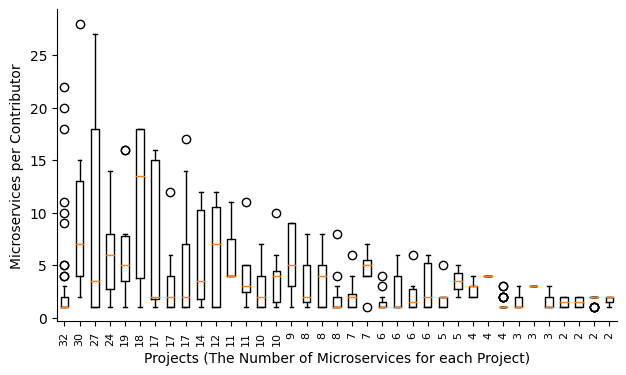}
  \caption{MS Per developer sorted by \#Microservice (RQ$_2$)}
  \label{fig:sortbyms}
\end{minipage}
\label{fig:papers}
\vspace{-4mm}
\end{figure*}

To tackle $RQ_2$, we first investigated the strategies of different contributor-microservice effort allocations. \Cref{fig:sortbyms} shows the distribution of the ``microservices per developer" of each of the selected projects. To be noted, the \textit{light-example-4j}, which contains 155 different microservices, is not shown in \Cref{fig:sortbyms}. Because one outlier in this project reaches 154 microservices, showing this project in the chart will make the details of all other projects invisible. Nonetheless, this project was certainly included in the analysis process.


From \Cref{fig:sortbyms}, we can easily find that for all the projects, the \textit{one-microservice-per-developer} strategy has not been respected. For the selected projects, the majority of the medians range from one to seven. For all projects, there are always some developers committing across multiple microservices. 


On the contrary, many projects that contain various numbers of microservices have one individual contributor who contributes to all the microservices. We name such a strategy \textit{One-Dev-ALL-MS}. For example, in project \textit{geoserver-cloud}, contributor \texttt{gabriel.roldan} committed in all the 11 microservices, and in project \textit{eShopOnContainers}, contributor \texttt{mvelosop} covers all the 17 microservices. Furthermore, many projects even have multiple contributors that cover all the microservices. We name such a strategy \textit{Multi-Dev-All-MS}. For example, in project \textit{loopback4-microservice-catalog}, there are eight contributors covering all 18 microservices; and in project \textit{DeathStarBench} six contributors cover all three microservices. It is likely that such a phenomenon is irrelevant to either the microservice number or a number of contributors.

Based on this phenomenon, we can intuitively categorize the projects as follows.

\begin{itemize}
    \item \textit{One-Dev-ALL-MS projects}: Projects where only one individual contributor covers all microservices while all the others cover part of them (16 out of 38) 
    \item \textit{Multi-Dev-ALL-MS projects}: Projects with multiple contributors covering all microservices (10 out of 38)
    \item \textit{Multi-Dev-SOME-MS projects}: Any projects with no contributors covering all microservices; nor do they adopt ``One-microservice-per-developer'' strategy (12 out of 38)
    \item \textit{One-MS-per-developer projects}: Any projects with each contributor/team working only on one microservice
\end{itemize}

To further investigate the potential roles of the contributors that cover all microservices and the other common contributors, we used EFA to detect the latent factors. 

\textbf{1. Preprocessing}.
Firstly, for the preprocessing, we grouped the original dataset by the contributors. For each contributor, we synthesized his/her contribution in every language by checking the extensions of the committed files. We crawl each project's languages using GitHub API. By grouping the data, we obtained the 1\,536 contributors' dataset with their contribution to the 33 languages. And we further normalized the dataset into values between zero and one.




\textbf{2. Data Verification}. 
Herein, the KMO score for this dataset is 0.585. It shows that the sampling is adequate and applying factor analysis is useful for this dataset. When applying PA to the dataset, we detected 13 factors as there are 13 out of 33 observed eigenvalues greater than 1.0. The corresponding factor loadings are shown in the replication package\footref{Package}.


    
    

    
    
    


\textbf{3. Determining Factor Number}. Based on the result of the parallel analysis (PA), the turning point can be found easily by examining the differences between observed
eigenvalues and simulated eigenvalues. Since the simulated eigenvalue becomes greater than the observed eigenvalue in the 14th factor (1.00049 and 0.90517, respectively), the first 13 factors are retained. The number of factors is therefore 13. According to Guadagnoli and Velicer \cite{guadagnoli1988relation}, scores greater than 0.4 are considered stable, especially when all variables are not cross-loaded~heavily. 


\textbf{4. Factor Extraction and Interpretation}. The initially detected factors and the correlated variables are reported in the replication package\footref{Package}. 
Herein, we adapted Montandon et al.'s role-language relevance results \cite{montandon2021mining} as the reference to analyze the interpretation of each factor. To be noted, we added several languages that are not listed in Montandon et al's study based on common knowledge and experts' opinions. 

Meanwhile, we also considered the other contributors that are not related to any specific roles above as \textit{Others}. By calculating the KH similarity between the role factors in the factor table obtained previously and the reference table \cite{montandon2021mining}. Here we assigned the role with the highest similarity score to each factor. 


Furthermore, we combined the factors with the same roles and obtained the final role-factor reference model. 

\textbf{5. Determining Individual Developer Role Allocation}. By using this role-factor relevance model, we simply calculated the factor similarities of any contributor, given his/her contribution allocation in terms of the 33 languages. Furthermore, we investigated the difference in terms of the contributor roles of the project strategies mentioned above. 

\begin{figure}
    \centering
    \includegraphics[width=0.6\textwidth]{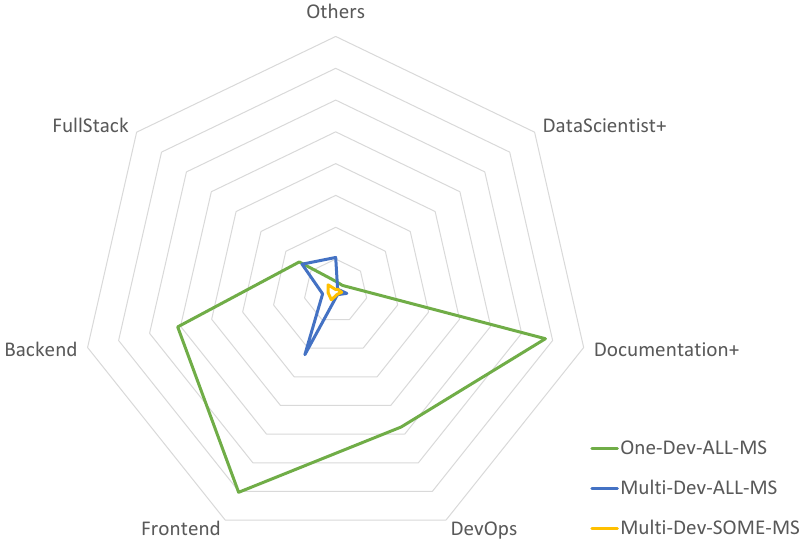}
    \caption{Average Role-Factor Distribution of Each Strategy (RQ$_2$)}
    \label{fig:strategyrolefactors}
\end{figure}

Figure \ref{fig:strategyrolefactors} shows the average behavior patterns of the different types of contributors in terms of the technical roles. From Figure \ref{fig:strategyrolefactors}, we can easily observe that the individual contributors who cover all microservices (i.e., One-Dev-ALL-MS) of the projects contribute largely as Documentation+. And they are also heavily involved in Frontend, when slightly less in Backend and DevOps roles. To be noted, they also contribute as Fullstack but are nearly non-existent in the other aspects. In addition, the One-Dev-ALL-MS also contributes as the Data Scientist role more than the others. On the other hand, for the multiple contributors that cover all the microservices (i.e., Multi-Dev-ALL-Ms), these contributors, on average, contribute less than the One-Dev-ALL-MS mentioned above. However, they contribute slightly more as Frontend than the other roles. They cover the Fullstack role a little less but surprisingly at a similar level compared to One-Dev-ALL-MS. Furthermore, they contribute more in other languages that are not role-related than that from the One-Dev-ALL-MS. The Multi-Dev-ALL-MS also contributes to Backend, Documentation, and Data Scientist, but much less than the other aspects. Regarding all the Multi-Dev-SOME-MS contributors, they contribute much less in terms of all working roles than the One-Dev-ALL-Ms and Multi-Dev-ALL-Ms. 

\keyfindings{
The majority of the microservice projects have one or multiple contributors who commit to all microservices. The single contributor who covers all microservices (One-Dev-ALL-MS) contributes much more than the multiple contributors covering all microservices (Multi-Dev-ALL-MS) in all roles, except that Multi-Dev-ALL-MS contribute more in non-role-related languages. Multi-Dev-SOME-MS contribute much less in all roles.}



\section{Discussion}
\label{sec:Discussion}

Using a large established 145-microservice project dataset \cite{DamianDataSet} we selected 38 projects with a sufficient number of contributors and commits as a representative OSS sample. This project sample did not adopt the same strategy suggested for proprietary (closed-source) software projects. In the analyzed sample we identified that developers typically work on multiple microservices, and focus on various features, often in parallel. These conclusions are also confirmed by the vast majority of the projects when considering different developer roles. 

One of the explanations might be the dynamics of OSS projects. In OSS projects, developers commit their time voluntarily at random, non-fixed hours and schedules, oftentimes driven by feature priority requests or error reports. In particular, none of the selected projects is directly sponsored by a company that allocates developers to the project. Therefore, developers commonly select a set of issues to be implemented (either new features or bug fixing) and work on them independently rather than adopting the specific microservice that they maintain.  
Another explanation might be that, despite the decentralized nature of OSS and the microservice architecture, OSS projects might not have yet assimilated this strategy. Another reason might be the lack of clear teams in OSS projects (i.e. each developer  does not belong to a specific team), and therefore the ``one-microservice-per-developer'' strategy might not be perceived as an issue. 

It must also be recognized that additional effort and overhead are related to the ``one-microservice-per-developer'' strategy. However, this might not be the proper fit for the OSS environment and context. OSS projects are often driven by small development teams or individuals who stand behind the entire project, occasionally OSS projects have professional teams behind them (i.e., Red Hat); however, we did not include these projects in the study. 

Microservice architecture is the mainstream architecture for cloud-native systems. However, not necessarily all microservice systems are cloud-native. In a similar parallel, the decentralized development model connected with cloud-native systems might collide with the OSS development model. Perhaps the main driver for the microservice architecture in these OSS projects is scalability and the decentralized development aspect goes away with the OSS model.

As practitioners often suggest~\cite{DamianDataSet,7436659,Amazon,microservice.io,qcon@2022.darkenergy}, if the development team is too small to be split into multiple teams, and there are multiple microservices, to respect the one-microservice-per-developer strategy, the system should rather remain monolithic. The reasons for OSS might be prioritized system scalability for the price of this strategy violation. 
Perhaps some projects might have decided to split their systems into multiple microservices for maintainability reasons, to increase the separation of concerns, or to better identify different business domains, independently from the team that is working on the same services. 

Another explanation might be given by Mariusz, who investigated whether Conway’s Law applies to OSS projects \cite{Mariusz2019} and concludes that teams "organize themselves spontaneously around tasks, and since those tasks concern software modules, teams naturally follow Conway’s law". 

The result of this study will serve the practitioners' community to understand how OSS microservice projects are being developed. Moreover, it will help researchers to further investigate the one-microservice-per-developer strategy.
\section{Threats to Validity}
\label{sec:ThreatsValidity}


\textit{Construct Validity.}  Replying to \textbf{RQ$_1$} we tried to understand if the \textit{one-microservice-per-developer} strategy is adopted. But we measure how well this strategy is adopted in an OSS context by analyzing individual developer behavior and assuming that a single developer belongs to one team at a time. We recognized that this assumption could lead to some threats.
We planned to expand our work in the future by adding information (such as developer communications, and issue/pull request comments) to extract teams to fine-grain our analysis.

\textit{Internal Validity.} The dataset used is one of the most recent in the context of microservices and open-source projects. 
However, the dataset is very heterogeneous (for the number of microservices, the age of the projects, and the number of developers), and we could only analyze a subset of the projects. 
We want to extend the dataset to get a better picture of the real state of the art.

\textit{External Validity.} 
The findings of this paper can be simply extended when more microservice projects are taken into account. It is reasonable that all the currently included projects shall also inevitably evolve when the proposed method should be replicated with the results updated. Especially for \textbf{RQ$_2$}, the findings can also be generalized to projects that are not specifically microservice-based if we use modules or features to functionally separate the projects instead of using microservices. In this way, such extended findings shall provide insights into the collective contributor profiles for any given OSS project scope. In addition, when the language-role relations can be further defined (e.g., new roles defined, new languages assigned to different roles, etc.), the findings can also be updated accordingly with the changes conducted in the reference table. 

\textit{Reliability.} Using the dataset we provided in the replication package\footref{Package} with the same approach, the practitioners and scholars can easily obtain the same results as described above. Only when any changes are introduced in the data itself or when the interpretation of the obtained factors varies based on different expertise, the findings shall differ accordingly. 

\section{Conclusion}
\label{sec:Conclusion}
Based on the suggestion of practitioners that "a developer should have exactly one service unless there is a proven need to have multiple services" and the assumption that developers developing open source software using a bazaar-style software development model would encourage a "one microservice per developer" strategy, we learned in this study that OSS projects do not comply with this strategy.
Oftentimes, we could identify projects with a greater number of microservices than project contributors, and the OSS development model with a few main contributors dominated the proprietary software strategy. Still, we must assume that the contributor dedication to OSS has a very different dynamics than fully-funded organization projects that can afford multiple developers with regular commitments to contribution. One might question if Conway's law collides with the OSS development model, and the results of this study add weight to the doubts. In this work, we showed that OSS microservice projects rarely follow the ``one-microservice-per-developer'' strategy. 

We have demonstrated this by analyzing the OSS project source code repositories of an established microservices project dataset. We further supported this result by analyzing the different developers' roles in contributing to these projects.

As future work, we aim at further study if ``one-microservice per developer" holds in OSS projects trying to observe emerging or stable developer-like collaborations between developers. To do so, we plan to analyze the commits of source code repositories of microservice projects, also parsing the actual code modifications to understand if a collaboration took place. Also, following the suggestion by Mariusz\cite{Mariusz2019} projects, which states that ``developers organize themselves spontaneously around tasks,''  we plan to study issue-tracking systems in combination with source code repositories to investigate if we are able to detect such spontaneous developers acting on single microservices.
Moreover, we aim to investigate the developers' team composition to classify the developers who contribute to the same code.

\bibliographystyle{splncs04}
\bibliography{sample-bibliography.bib}

\end{document}